\documentstyle [aps,psfig]{revtex}

\begin{document}
\draft
\title{Evidence for an oscillatory singularity in generic $U(1)$
symmetric cosmologies on $T^3 \times R$} 
\author{Beverly K. Berger \thanks{
e-mail address: berger@oakland.edu}}
\address{Department of Physics, Oakland University, Rochester, MI 48309 USA}
\author{Vincent Moncrief  \thanks{
e-mail address: moncrief@hepvms.physics.yale.edu}}
\address{Departments of Physics and Mathematics, Yale University, New Haven, CT
06520 USA}

\maketitle
\bigskip
\begin{abstract}
A longstanding conjecture by Belinskii, Lifshitz, and Khalatnikov that the singularity in
generic gravitational collapse is locally oscillatory is tested numerically in vacuum,
$U(1)$ symmetric cosmological spacetimes on $T^3 \times R$. If the velocity term
dominated (VTD) solution to Einstein's equations is substituted into the Hamiltonian for
the full Einstein evolution equations, one term is found to grow exponentially. This
generates a prediction that oscillatory behavior involving this term and another (which
the VTD solution causes to decay exponentially) should be observed in the approach to the
singularity. Numerical simulations strongly support this prediction.

\end{abstract}

\section{Introduction}
An important open question in classical general relativity is the nature of the
singularities that form in generic gravitational collapse. The singularity theorems of
Penrose \cite{penrose69}, Hawking \cite{hawking67,hawking70}, and others
\cite{hawking73,wald84} prove that some type of singular behavior must arise in generic
gravitational collapse of reasonable matter. However, these theorems do not provide a
description of the singular behavior that results. Many different types of singular
behavior arise in known solutions to Einstein's equations. However, known explicit
solutions tend to be characterized by simplifying symmetries. This means that singular
behavior found in such examples need not be characteristic of those of generic collapse.
In the 1960's, Belinskii, Khalatnikov, and Lifshitz (BKL)
\cite{belinskii69a,belinskii69b,belinskii71a,belinskii71b,belinskii82} claimed to have
shown that, in the approach to the singularity, each spatial point of a generic solution
behaves as a separate vacuum, Bianchi IX (Mixmaster
\cite{misner69}) homogeneous cosmology. Mixmaster cosmologies collapse as an infinite
sequence of Bianchi I (Kasner \cite{kasner25}) spacetimes with a known relationship
between one Kasner and the next
\cite{belinskii71b,chernoff83}. Examinations of the BKL arguments \cite{barrow79} and
attempts to provide a more rigorous basis for their claims \cite{grubisic94a} have until
recently yielded little evidence one way or the other for their validity.

In the past decade, work has begun aimed at understanding the singularity and the approach
to it in spatially inhomogeneous cosmologies
\cite{isenberg90,grubisic93,berger93}. If Einstein's equations are truncated by ignoring
all terms with spatial derivatives and keeping all terms with time derivatives (for some
more or less natural choice of spacetime slicing), the velocity term dominated (VTD)
solutions are found. If an inhomogeneous spacetime is asymptotically VTD (AVTD), the
evolution toward the singularity at (almost) every spatial point comes arbitrarily close
to one of the VTD solutions \cite{isenberg90,eardley72}. It has been proven that polarized
Gowdy cosmologies
\cite{gowdy71,berger74} and more general polarized $T^2$-symmetric models \cite{berger97o}
are AVTD \cite{isenberg90,isenberg98}. The main extra feature in generic Gowdy models
compared to polarized ones is the presence of two nonlinear terms in the Hamiltonian which
yields the dynamical Einstein equations. In an AVTD spacetime, these must become
exponentially small as the singularity is approached. But this requrement is only
consistent with the form of the VTD solution if a spatially dependent parameter of the VTD
solution lies in a restricted range at (almost) every spatial point. The numerical studies
demonstrate that the nonlinear terms act as potentials to drive the parameter into the
consistent range
\cite{berger97b}. Very recently, it has  been proven that generic Gowdy solutions with a
consistent value of this parameter are AVTD \cite{kichenassamy98}. 

Recently, Weaver et al \cite{weaver98} have extended the Gowdy model by inclusion of a
magnetic field (and change of spatial topology from $T^3$ to the solv-twisted torus
\cite{fujiwara93}). This model is an inhomogeneous generalization of the magnetic Bianchi
VI$_0$ homogeneous cosmology which is known to exhibit Mixmaster behavior
\cite{leblanc95,berger96a}. The magnetic field causes a third nonlinear term to be
present. This term grows for precisely that range of VTD parameter which makes the
original two vacuum Gowdy nonlinear terms exponentially small. This prediction of local
Mixmaster oscillations is observed in numerical simulations of the full Einstein
equations. This study provided the first support in inhomogeneous cosmologies for the BKL
claim.

To further explore this issue, we have generalized to the study of vacuum spacetimes on
$T^3 \times R$ with one spatial $U(1)$ symmetry \cite{moncrief86}. While such models are,
of course, not generic solutions to Einstein's equations, they are considerably more
complex than any previously considered for this purpose. Application of the methods
previously described for Gowdy models leads to the predictions that (1) {\it polarized}
$U(1)$ symmetric cosmologies should be AVTD \cite{grubisic94} and (2) generic $U(1)$
models should have an oscillatory singularity. Numerical simulations have previously
provided strong support for the predicted AVTD singularity in polarized $U(1)$ models
\cite{berger97e}. Here we shall discuss the support we have obtained for the local
oscillatory nature of the singularity in generic $U(1)$ models.

Unlike all previous cases discussed in this program
\cite{berger93,berger97b,weaver98,berger97e}, numerical difficulties require an
introduction of spatial averaging (data smoothing) at each time step to prevent numerical
instability. This averaging destroys convergence of the solution with increasing spatial
resolution (see \cite{choptuik91}). The need for spatial averaging can be traced to the
growth of spiky features first seen and discussed in the Gowdy models
\cite{berger93,berger97b}. These features also (as the bounces in Mixmaster itself
\cite{berger96c}) make it necessary to explicitly enforce the Hamiltonian constraint. If
this is not done, qualitatively incorrect behavior will result. We shall argue in this
paper that, despite the need for spatial averaging and our simple algebraic method of
enforcement of only the Hamiltonian constraint, the qualitative behavior of the numerical
simulations is correct. We believe we have demonstrated that the singularity in generic
$U(1)$ symmetric cosmologies is spacelike, local, and oscillatory. Evidence for this is
that the oscillations and their correlation with the exponential growth of certain
nonlinear terms in Einstein's equations are independent of spatial resolution and choice
of initial data.

A brief review of generic $U(1)$ models, their VTD solution, and the prediction of
oscillatory behavior are given in Section II. In Section III we describe the numerical
issues and results. Discussion is given in Section IV.

\section{The Model}
As discussed in more detail elsewhere \cite{berger97e}, $U(1)$ symmetric cosmologies on
$T^3
\times R$ are described by the metric
\begin{equation}
\label{metric}
ds^2=e^{-2\varphi }\left\{ {-e^{2\Lambda}e^{-4\tau }d\tau ^2+e^{-2\tau }e^\Lambda
e_{ab}dx^adx^b} \right\}+e^{2\varphi }\left( {dx^3+\beta _adx^a} \right)^2
\end{equation}
where $\varphi$, $\beta_a$, $z$, $x$, $\Lambda$ are
functions of spatial variables $u$, $v$ and time $\tau$,  sums are over $a,\,b =
u,\,v$, and 
\begin{equation}
\label{eab}
e_{ab}={\textstyle{{1 \over 2}}}\left[
{\matrix{{e^{2z}+e^{-2z}(1+x)^2}&{e^{2z}+e^{-2z}(x^2-1)}\cr
{e^{2z}+e^{-2z}(x^2-1)}&{e^{2z}+e^{-2z}(1-x)^2}\cr }} \right]
\end{equation}
is the conformal metric of the $u$-$v$ plane. 
A canonical transformation replaces the twists $\beta_a$ and their conjugate
momenta $e^a$ with the twist potential $\omega$ and its conjugate momentum $r$
\cite{moncrief86,berger97e}. The dynamical variables $\varphi$ and $\omega$ are
respectively related to the amplitude for the
$+$ and $\times$ polarizations of gravitational waves and propagate in a background
spacetime described by $z$, $x$, and $\Lambda$. Our coordinate choice ($N = e^\Lambda$,
zero shift) does not significantly restrict the generality of these models
\cite{moncrief86}. Einstein's evolution equations (in vacuum) are found from the variation
of \cite{berger97e} 
\begin{eqnarray}
\label{Hu1}
H &=& \int \int du \kern 1pt dv \,{\cal H} \nonumber \\
&=& \int \int du \kern 1pt dv \left( {\textstyle{{1 \over 8}}}p_z^2+{\textstyle{{1 \over
2}}} e^{4z}p_x^2+{\textstyle{{1 \over 8}}}p^2+{\textstyle{{1 \over 2}}}e^{4\varphi
}r^2-{\textstyle{{1 \over 2}}}p_\Lambda ^2+2p_\Lambda  \right) \nonumber \\
&& +e^{-2\tau } \int \int du \kern 1pt 
dv \left\{  \left( {e^\Lambda e^{ab}} \right) ,_{ab}- \left( {e^\Lambda e^{ab}}
\right) ,_a\Lambda ,_b+e^\Lambda  \right. \left[  \left( {e^{-2z}}
\right) ,_u x,_v- \left( {e^{-2z}} \right) ,_v x,_u \right] \nonumber \\
&& \left. +2e^\Lambda e^{ab}\varphi ,_a\varphi ,_b+{1 \over 2}
e^\Lambda e^{-4\varphi }e^{ab}\omega ,_a\omega ,_b \right\} \nonumber \\
&=& H_K+H_V = \int \int du \kern 1pt dv \, {\cal H}_K +\int \int du \kern 1pt dv
\,{\cal H}_V 
\end{eqnarray}
where \{$p$, $r$, $p_x$, $p_z$, $p_\Lambda$\} are respectively canonically conjugate to
\{$\varphi$, $\omega$, $x$, $z$, $\Lambda$\}.  The constraints are 
\begin{equation}
\label{H0}
{\cal H}^0 = {\cal H} - 2 p_\Lambda = 0
\end{equation}
and
\begin{eqnarray}
\label{Hu}
{\cal H}_u&=&p_z\,z,_u+p_x\,x,_u+p_\Lambda \,\Lambda ,_u-p_\Lambda ,_u+p\varphi
,_u+r\omega ,_u+{\textstyle{{1 \over 2}}}\left\{ {\left[ {e^{4z}-(1+x)^2}
\right]p_x-(1+x)p_z} \right\},_v \nonumber \\
  & &-{\textstyle{{1 \over 2}}}\left\{ {\left[ {e^{4z}+(1-x^2)} \right]p_x-x\kern 1pt p_z}
\right\},_u=0,
\end{eqnarray}
\begin{eqnarray}
\label{Hv}
{\cal H}_v&=&p_z\,z,_v+p_x\,x,_v+p_\Lambda \,\Lambda ,_v-p_\Lambda ,_v+p\varphi
,_v+r\omega ,_v-{\textstyle{{1 \over 2}}}\left\{ {\left[ {e^{4z}-(1-x)^2}
\right]p_x+(1-x)p_z} \right\},_u \nonumber \\
 & &+{\textstyle{{1 \over 2}}}\left\{ {\left[ {e^{4z}+(1-x^2)} \right]p_x-x\kern 1pt p_z}
\right\},_v=0.
\end{eqnarray}
The VTD solution has been given elsewhere \cite{berger97e}. Here we shall consider only
the limit as $\tau \to \infty$ of this solution. (Recall that the VTD solution is found by
eliminating all terms with spatial derivatives from Einstein's equations.) The limiting
VTD solution is
\begin{eqnarray}
\label{u1avtd} 
z&=&-v_z\tau,  \quad, x= x_0,  \quad p_z= -4v_z, \quad
p_x= p_x^0, \quad
\varphi= -v_\varphi\tau, \nonumber  \\
\omega&=& \omega_0, \quad
p=-4v_\varphi,  \quad
r= r^0, \quad
\Lambda = \Lambda_0 +(2 - v_\Lambda)\tau, \quad
p_\Lambda = v_\Lambda
\end{eqnarray}
where $v_z$, $v_\varphi$, $x_0$, $p_x^0$, $\omega_0$, $r^0$, $\Lambda_0$, and
$v_\Lambda > 0$ are functions of $u$ and $v$ but independent of $\tau$. (The sign of  
$v_\Lambda$ is fixed to ensure collapse.)

We now use the method of consistent potentials (MCP) \cite{grubisic93} to determine the
consistency of the VTD solution with the full Einstein equations. As $\tau \to \infty$,
consistency requires that all terms other than those which yield (\ref{u1avtd}) should
become exponentially small. Rather than consider the equations, we shall examine the
Hamiltonian density which generates them. Possible inconsistencies could arise from the
nonlinear terms in (\ref{Hu1}) containing exponential factors. These same exponentials
would, of course, be present in Einstein's equations. We first notice that the Gowdy-like
terms \cite{berger93}
\begin{equation}
\label{u1v1}
V_z = {\textstyle{1 \over 2}}\, p_x^2 \,e^{4z}, \quad V_1 = {\textstyle{1
\over 2}} \,r^2 \,e^{4\varphi } 
\end{equation}
in ${\cal H}_K$ become, in the limit of $\tau \to \infty$, upon substitution of
(\ref{u1avtd})
\begin{equation}
\label{u1v1lim}
V_z  \to {\textstyle{1 \over
2}}\, p_x^2 \,e^{-4v_z \tau}, \quad V_1  \to {\textstyle{1
\over 2}} \,r^2 \,e^{-4v_\varphi \tau}
\end{equation}
and are exponentially small only if $v_z > 0$ and $v_\varphi > 0$. (As in the Gowdy case
\cite{berger97b}, non-generic behavior can arise at isolated spatial points where $p_x$
and/or
$r$ vanish.)

The complicated terms in $H$ containing the spatial derivatives have only two types of
exponential behavior. All but one of the terms in ${\cal H}_V$ have a factor
\begin{equation}
\label{u1vlam}
e^{(-2 \tau + \Lambda - 2z)}
\end{equation}
(if we assume $v_z > 0$, all components of $e_{ab}$ are dominated by $e^{-2z}$) which
becomes
\begin{equation}
\label{u1vlamlim}
 \approx e^{(-v_\Lambda +2 v_z) \tau}
\end{equation}
in the VTD limit. The remaining term is 
\begin{equation}
\label{u1v2}
V_2 = {\textstyle{1 \over 2}} e^{-2 \tau + \Lambda}e^{-4\varphi}e^{ab}\omega,_a \omega_b
\end{equation}
which becomes upon substitution of (\ref{u1avtd})
\begin{equation}
\label{u1v2lim}
V_2 \approx F(x,\nabla \omega) e^{(-v_\Lambda + 2 v_z + 4 v_\varphi) \tau}
\end{equation}
where $F$ is some function. The coefficients of $\tau$
in (\ref{u1vlamlim}) and (\ref{u1v2lim}) are restricted by the VTD form of the
Hamiltonian constraint (as $\tau \to \infty$)
\begin{equation}
\label{u1havtd}
{\cal H}^0 \approx -{\textstyle{1 \over 2}} v_\Lambda^2 +2 v_z^2 + 2
v_\varphi^2 \approx 0
\end{equation}
obtained by substitution of (\ref{u1avtd}) into (\ref{H0}).
As discussed in \cite{berger97e}, (\ref{u1havtd}) implies that $v_\Lambda > 2 v_z$ so that
(\ref{u1vlamlim}) decays exponentially for $v_z > 0$ for any $v_\varphi$. On the other
hand, for
$V_2$ to become exponentially small with $v_z$ and $v_\varphi$ $>0$, we require
$v_\Lambda^2 > (2 v_z + 4 v_\varphi)^2$ which is inconsistent with (\ref{u1havtd}). Since
there is no way to make $V_1$ and $V_2$ both exponentially small with the same value of
$v_\varphi$, the MCP predicts that either $V_1$ or $V_2$ will always grow exponentially.
(Again, non-generic behavior can result at isolated spatial points where the coefficient
of $V_2$ happens to vanish.)

To refine this prediction, consider $v_z >> |v_\varphi|$. Substitution in (\ref{u1havtd})
yields
\begin{equation}
\label{approxvl}
v_\Lambda \approx 2 v_z + {{v_\varphi^2} \over {v_z}}
\end{equation}
which shows that
\begin{equation}
\label{vphidep}
V_2 \approx e^{4 v_\varphi \tau} F(x,\nabla \omega)\ \ , \quad V_1 \approx r^2 e^{-4
v_\varphi \tau}.
\end{equation}
Thus we expect $V_1$ and $V_2$ to act as potentials for the $\varphi$ degree of freedom
with a bounce off either potential causing the sign of $v_\varphi$ to change. The
remaining variables will follow the VTD solution with parameters which change at every
bounce in $\varphi$. Thus we predict that oscillations in the $\varphi$ degree of freedom
will occur at (almost) every spatial point with different values of $v_\varphi$ and
coefficients of $V_1$ and $V_2$.

Note that in polarized $U(1)$ models, $\omega = r = 0$ so that the oscillations of
(\ref{vphidep}) should be absent. Polarized $U(1)$ models should thus be AVTD. This is
precisely what has been found in numerical simulations of these models \cite{berger97e}.

\section{Results and Numerical Issues}
In order to test the predictions of the previous section, we performed numerical
simulations of the Einstein evolution equations obtained from the variation of
(\ref{Hu1}). We use a symplectic PDE solver which has been described in great detail
elsewhere
\cite{berger93,berger97e,fleck76,moncrief83,suzuki90,suzuki91,norton92}. In our previous
study of polarized
$U(1)$ models, we demonstrated convergence of the solutions at the expected order with
increasing spatial resolution. Unfortunately, the re-introduction of the $\omega$ degree
of freedom leads to the growth of spiky features absent in the polarized case. The origin
of the spiky features is discussed elsewhere \cite{berger97b} but is related to the
non-generic behavior at isolated spatial points. The spiky features which our methods
easily treat in one spatial dimension \cite{berger93} cannot be modeled sufficiently well
in two spatial dimensions to prevent numerical instability. However, these instabilities
can be suppressed in two ways. First, the simulations are begun at $\tau \approx 10$ or
so to reduce the influence of ${\cal H}_V$ from (\ref{Hu1}). (This has the disadvantage,
as in homogeneous Mixmaster spacetimes \cite{khalatnikov85}, of increasing the time
interval between bounces.) Second, at every time step, all ten variables are replaced by
their spatial ``average.'' Any function $f(u,v)$ is replaced by \cite{norton92}
\begin{eqnarray}
\label{average}
\bar f(u,v) &=& c_0 f(u,v) + \sum_{i=1}^5 \, c_i
\,[f(u+i\Delta u,v )+f(u,v + i\Delta
v) \nonumber \\
& &+f(u-i\Delta u,v)+f(u,v- i\Delta
v)-4f(u,v)]
\end{eqnarray}
where $c_0 = 1/2 $, $c_1 = 4867/38400$, $c_2 = -1067/28800$, $c_3 = -1237/691200$, $c_4 =
787/345600$, $c_5 = 31/691200$ and $\Delta u$, $\Delta v$ are the grid spacings. This
scheme is 6th order accurate (i.e. the difference between $\bar f(u,v)$ and $f(u,v)$ is
7th order in the grid spacings---actually 8th order due to symmetry). Where $f(u,v)$ is
smooth, the averaging has no effect. However, it does spread out grid scale size spiky
features. Unfortunately, the averaging process destroys the convergence seen in the
polarized case (by examination of the deviation of the Hamiltonian and momentum
constraints from zero
\cite{choptuik91}), as can be determined by adding averaging to polarized $U(1)$
simulations. We shall argue later that despite the absence of convergence in this sense,
the qualitative behavior is still independent of spatial resolution. We shall further
argue that the correct qualitative behavior is sufficent to determine the nature of the
singularity in generic
$U(1)$ models.

In Section II, we showed that the behavior of the potentials in ${\cal H}_V$ was
restricted by the VTD limit of the Hamiltonian constraint (\ref{u1havtd}). This means that
it is essential to preserve the Hamiltonian constraint during the simulation. Failure to
preserve ${\cal H}^0 = 0$ means that $p_\Lambda$ would have the {\it wrong} dependence on
$p_z$ and $p$. This could change the sign of the coefficient of $\tau$ in (\ref{u1v2lim}).
While solving the constraints initially is sufficient analytically, there is no way to
guarantee that the differenced form of the constraints is preserved during a numerical
simulation of the Einstein evolution equations. While in the polarized
$U(1)$ case, the constraints can be seen to converge to zero \cite{berger97e}, this is no
longer true in generic $U(1)$ simulations with spatial averaging. To preserve the correct
relationships among
$v_\Lambda$, $v_z$, and $v_\varphi$ from (\ref{u1havtd}), we therefore solve ${\cal H}^0 =
0$ (\ref{H0}) for
$p_\Lambda > 0$ algebraically at each time step. We note that this is the precise analog
of the procedure used in Mixmaster itself \cite{berger96c}. However, spatial averaging
must be performed on this new $p_\Lambda$ which then yields a (small) non-zero value for
${\cal H}^0$. If we assume that $p_\Lambda = p_\Lambda^{\rm true} + \Delta$ is the
measured value, then (\ref{u1havtd}) can be linearized in the error $\Delta$ and solved
to yield
\begin{equation}
\label{delta}
\Delta = - {{{\cal H}^0} \over {p_\Lambda}}
\end{equation}
where ${\cal H}^0$ is the measured, erroneous value of the Hamiltonian constraint.
Substitution in $\alpha = - v_\Lambda +2 v_z +4 v_\varphi$, the coefficient of $\tau$ in
$V_2$, yields a measured and true value for $\alpha$. If the product $\alpha \,
\alpha^{\rm true}$ is positive everywhere for all $\tau$, the errors due to averaging
cannot have changed the qualitative behavior since that just depends on the sign of
$\alpha$. Examination of the simulation data shows $\alpha \, \alpha^{\rm true} > 0$ at
(almost) all spacetime points. Occasionally, early in the simulation,
a few isolated points---in space and time---will show a sign change due to this error.

The momentum constraints (\ref{Hu}) and (\ref{Hv}) are freely evolved and do not remain
especially small. However, they contain only spatial derivatives. Errors in the momentum
constraints would therefore generate errors in the spatial dependence  of the variables
at a given time but not in their qualitative time dependence at each spatial point. A
measure of the constraint convergence vs $\tau$ is shown in Figure 1. The freely evolved
${\cal H}_u$ (${\cal H}_v \approx {\cal H}_u$) is actually larger at finer spatial
resolution. This may be attributed to the larger spatial gradients observed at finer
spatial resolution
\cite{berger97b}. On the other hand, the error in the Hamiltonian constraint is converging
to zero.

The restricted initial value solution for generic $U(1)$ models has been described
elsewhere \cite{berger97e}. To solve the momentum constraints, we set $p_z = p_x =
\varphi,_a = \omega,_a = 0$ and $p_\Lambda = c e^\Lambda$. For $c > 0$ sufficiently large,
this allows the Hamiltonian constraint to be solved algebraically for $p$ (case A) or $r$
(case B). Four functions, $x$, $z$, $\Lambda$, and $r$ (case A) or $p$ (case B), may be
freely specified. The detailed values used in the numerical simulations are given in the
figure captions. While we shall present results for only two sets of initial data, the
results are typical of all initial data and may be regarded to be characteristic of
generic initial data.

Figures 2-4 show the behavior of $\varphi$, $V_1$, and $V_2$ for three spatial
resolutions for case A initial data. In all cases, the behavior is exactly as predicted
in Section II. Figure 5 shows a typical evolution for case B initial data, again showing
the predicted behavior. For the simulation of Figure 3, the remaining variables $\omega$,
$z$, $x$, and $\Lambda$ vs $\tau$ are shown at the same spatial point in Figure 6. Here we
see approximate VTD behavior in $\omega$, $z$, and $\Lambda$ with parameters changing at
the times of the
$\varphi$ bounces. While the behavior of $x$ does not appear to be as predicted, we note
from (\ref{u1avtd}), (\ref{u1vlam}), and (\ref{u1v2}) that only the behaviors of
$z$ and $\Lambda$ are important to the local dynamics since only they appear in
the arguments of exponentials. Finally, we display ``movie frames'' of
${\cal H}_V(u,v)$ and $\varphi(u,v)$ vs
$\tau$ in Figures 7 and 8. Since ${\cal H}_V$ is shown on a logarithmic scale and
$\varphi$ on a linear one, Figure 7 is dominated by $V_2$ while Figure 8 indicates the
behavior on a logarithmic scale of $V_1$. The predicted behavior that either $V_1$ 
or $V_2$ but not both be large is clearly seen in Figures 7 and 8. The bounces in
$\varphi$ occur at different spatial points at different times. This eventually will lead
to increasingly complex spatial structure
\cite{kirillov87}.

\section{Discussion}
Numerical simulations of generic $U(1)$ models demonstrate that the evolution toward the
singularity is local and oscillatory due to the alternate expoential growth and decay of
$V_1$ and $V_2$. This behavior requires that the time dependence of the coefficients of
$V_1$ and $V_2$ and the coefficients of $\tau$ in the approximate forms of $V_1$ and $V_2$
be negligible (i.e.~act on a much slower time scale) compared to the exponential time
dependence obtained from the VTD solution. The qualitative behavior seen in the
simulations indicates that this requirement is met (almost) everywhere sufficiently close
to the singularity. Exceptional behavior at isolated spatial points similar to that found
in simpler models \cite{berger97b,weaver98} cannot be studied with the current numerical
code. Exceptional behavior arises at isolated spatial points because one of the
exponential potentials which causes the generic behavior is absent. While we believe we
have identified the correct exponential behavior, the data we have on spatial dependence
at a given time is probably not sufficiently reliable for conclusions about ``higher
order'' effects.

The explanation for the local nature of the evolution is easy to obtain. From the metric
(\ref{metric}), we see that the distance $\Delta l$ traveled by a light ray away from the
singularity from $\tau = \infty$ to $\tau = \tau_0$ (coordinate horizon size) is
\begin{equation}
\label{horizon}
\Delta l \approx \int_{\infty}^{\tau_0}\, e^{\Lambda/2+z-\tau} d \tau
\end{equation}
where the VTD limit of $e_{ab}$ has been used. Simulations demonstrate that the VTD
solution may be used in (\ref{horizon}) to give $\Delta l \to 0$ for the coordinate
horizon size \cite{berger74} as $\tau_0 \to \infty$. Since the horizon size is decreasing,
the spatial points are unable to communicate with each other. But such communication
occurs through changing spatial derivatives. If no communication can occur, the spatial
derivative containing terms must be dynamically unimportant. 

The primary question then is the extent to which the numerical results presented here are
believable as evidence for the oscillatory nature of the singularity. It is easy to show,
e.g.~by comparing polarized $U(1)$ simulations with and without spatial averaging, that
spatial averaging ruins convergence tests. Furthermore, non-zero values of the momentum
constraints (\ref{Hu}) and (\ref{Hv}) indicate the presence of errors in the variables
and their spatial derivatives. (Since rescaling the coordinates by a constant
yields rescaled values of the constraints without changing the behavior of the
solutions, one cannot attach any significance to the actual magnitude of the
constraint violation.) Both sources of error---spatial averaging and constraint
violation---principally affect details of the spatial dependence of the variables at a
given time. Comparison of simulations at different spatial resolutions (and with quite
different initial data) shows that the qualitative time dependence at fixed spatial
points is not affected by these errors. This is shown in Figure 9 where $\varphi$ is
shown on a line $u = v$ with $u,v \in [0,\pi]$ vs $\tau$ for three different spatial
resolutions. While the features appear narrower at finer spatial resolution (see
\cite{berger97b} for a similar phenomenon in Gowdy models), the time development of the
simulations is remarkably consistent and independent of resolution. Figure 10 shows the
final time step for $\varphi$ for two different spatial resolutions. While the spatial
dependence is quantitatively different, it is qualitatively quite similar. We also see in
Figures 11 and 12, that the correlation between $V_1$ and $V_2$ seen at single spatial
points in Figures 2-6 occurs everywhere. Figure 11 shows $\log_{10}V_1$ and
$\log_{10}V_2$ on the line $u = v$ with $u,v \in [0,2\pi]$ vs $\tau$. Clearly, one
potential grows as the other decays at all spatial points. Figure 12 shows the final time
step for $V_1$, $V_2$, and
$\varphi$ on the line $u = v$ with $u,v \in [0,\pi]$. Given this same oscillatory behavior
in all the simulations and the explanation for it in terms of the VTD solution given in
Section II, it is proabable that we have correctly described the nature of the generic
singularity in this class of models.

This does not mean that it is unimportant to improve the numerical treatment. Better
numerics will yield more accurate spatial dependence of the metric variables. We expect
to see ever narrowing spatial structure as in Gowdy \cite{berger97b} and magnetic Gowdy
\cite{weaver98} models caused by non-generic values at isolated spatial points. This can
also yield precise data for analysis of the local dynamics to characterize the ``higher
order'' effects due to spatial inhomogeneity. It might also become possible to observe and
characterize any other exceptional behavior that might arise at isolated spatial points.

There are several possibilities for numerical improvement. First among these is to
incorporate into the algorithm the fact that a known explicit solution exists for the
bounce (scattering) off an exponential potential. The symplectic
algorithm divides the Hamiltonian for a system into two or more subhamiltonians with
explicit exact solutions for their equations of motion. The standard division of $H$ is
into kinetic ($H_K$) and potential ($H_V$) pieces. Including a dominant exponential wall
from $H_V$ in
$H_K$ has yielded tens of orders of magnitude of improvement in speed while maintaining
machine level precision in the ODE'S of spatially homogeneous Mixmaster models
\cite{berger96c}. The current $U(1)$ code uses ${\cal H}_K$ and ${\cal H}_V$ (\ref{Hu1})
as the subhamiltonians. However, one could treat
\begin{equation}
\label{hphi1}
{\cal H}_\varphi^{(1)} = {\textstyle{{1} \over {8}}} \, p^2 + {\textstyle{{1} \over
{2}}}e^\Lambda e^{ab} e^{4
\varphi} \omega,_a \omega,_b
\end{equation}
or
\begin{equation}
\label{hphi2}
{\cal H}_\varphi^{(2)} = {\textstyle{{1} \over {8}}} \, p^2 + {\textstyle{{1} \over
{2}}}r^2 \, e^{4 \varphi}
\end{equation}
as a separate subhamiltonian depending on which of $V_1$, $V_2$ is the larger. The
absence of $p_\Lambda$, $p_z$, and $r$ from (\ref{hphi1}) means that the equations from
${\cal H}_\varphi^{(1)}$ are exactly solvable (as, obviously are those from ${\cal
H}_\varphi^{(2)}$). While in the ODE case, the main advantage was found in the ability to
take huge time steps, for the $U(1)$ case, it will be the improved accuracy of the bounce
solution. Another area of improvement is in spatial differencing. The current code uses
4th and 6th order accurate representations of first and second derivatives due to Norton
\cite{norton92}. These are designed to minimize the growth of grid scale wavelength
instabilities. However, when applied to the Gowdy model, this scheme appears to be less
stable than the original one \cite{berger93} based on variation of a differenced form of
the Hamiltonian. We also note that spectral
methods have been tried but have not proven to be useful.
A third place for numerical improvement would be to solve all three constraints rather
than just the Hamiltonian constraint. 
Of course, one naturally wonders if adaptive mesh refinement (AMR) might yield
improvements in accuracy and stability. Studies to date have shown that increasing
spatial resolution yields only small improvements in stability. This is probably due to
the fact that finer resolution only gives better representation of spiky
features---showing them to be narrower and steeper than they appear at coarser resolution.
In the Gowdy simulations, however, it was noticed that, for any given simulation, there is
a threshold spatial resolution. If the resolution is coarser than that, the code will blow
up before the AVTD regime is reached everywhere. It is therefore possible that we have
not yet reached this threshold in the $U(1)$ case and thus would be helped by AMR.
However, in the generic
$U(1)$ models, there is no AVTD regime. Thus it is not clear if any spatial resolution
could yield a simulation which could run indefinitely without crashing.

As a final note, we remark that the local nature of the $U(1)$ evolution (as well as the
other cases we have studied) has two major implications:

(1) The use of cosmological boundary conditions is merely a convenience since they do not
affect the local behavior. Any collapse of a system with one spatial Killing field to a
spacelike singularity should be local and oscillatory.

(2) Qualitative answers on the nature of the singularity do not require fine spatial
resolutions. This means that the zero Killing field case should be tractable numerically.
Work on this case is in progress.

\section*{Acknowledgements}
We would like to thank the Albert Einstein Institute at Potsdam for
hospitality. BKB would also like to thank the Institute for Geophysics and Planetary
Physics of Lawrence Livermore National Laboratory for hospitality. This work was supported
in part by National Science Foundation Grants PHY9507313 and PHY9503133. Numerical
simulations were performed at the National Center for Supercomputing Applications
(University of Illinois).

\vfill
\eject

\section*{Figure Captions}
\bigskip
Figure 1. Convergence testing of constraints. The average values of the momentum
constraint ${\cal H}_u$ (broken line) and Hamiltonian constraint ${\cal H}^0$ (solid
line) are displayed vs $\tau$ for generic vacuum $U(1)$ symmetric model simulations with
$256^2$ (nothing) and
$512^2$ (triangles) spatial grid points. (See Figure 2 for initial data.)

\bigskip

Figure 2. Oscillatory behavior at a typical spatial grid point. The potentials $V_1$ and
$V_2$ and $\varphi$ are shown vs $\tau$ for a simulation with $128^2$ spatial grid
points. The initial data are $\Lambda = \sin u \, \sin v$, $x = z = \cos u \,  \cos v$,
$\varphi = \omega = 0$, $p_\Lambda = 14\, e^\Lambda$, $r = 10 \cos u \,  \cos v$, and $p_z
= p_x = 0$. The Hamiltonian constraint is solved for $p$. The initial value of $\tau $ is
$10$.

\bigskip

Figure 3. Oscillatory behavior at a typical spatial grid point. The same initial data as
in Figure 2 but with $256^2$ spatial grid points.

\bigskip

Figure 4.  Oscillatory behavior at a typical spatial grid point. The same initial data as
in Figure 2 but with $512^2$ spatial grid points.

\bigskip

Figure 5. Oscillatory behavior at a typical spatial grid point. The same as Figure 2 but
with Case B initial data:  $\Lambda = \sin u \, \sin v$, $x = z = \cos u \, \cos v$,
$\varphi = \omega = 0$, $p_\Lambda = 4\, e^\Lambda$, $p =  \cos u \, \cos v$, and $p_z =
p_x = 0$. The Hamiltonian constraint is solved for $r$. The initial value of $\tau $ is
$10$.

\bigskip

Figure 6. Behavior of (a) $\Lambda$ and $z$, (b) $\omega$, and (c) $x$ in the simulation
of Figure 3.

\bigskip

Figure 7. Movie frames of $\log_{10}{\cal H}_V(u,v)$ at (from right to left and top to
bottom) $\tau = 14.9$, $19.8$, $24.7$, $29.6$, $34.5$, $39.5$, $44.4$, $49.3$, $54.2$. The
range of values is $-30$ (black) to $3$ (white). The simulation of Figure
3 provided the data.

\bigskip

Figure 8. Movie frames of $\varphi(u,v)$ at the same values of $\tau$ as in Figure 7,
taken from the same simulation. The
range of value is $-27$ (black) to $3.8$ (white) on a linear scale.

\bigskip

Figure 9. The influence of spatial resolution. The variable $\varphi$ on the line $u =
v$ for $u,v \in [0,\pi]$ (vertical axis) is shown for $\tau \in [10,53.5]$ (horizontal
axis). The simulations of Figures 2 (top), 3, and 4 (bottom) are shown. The gray scale is
similar to that in Figure 8.

\bigskip

Figure 10. The variable $\varphi$ is shown for the line $u =
v$ for $u,v \in [0,\pi]$ at $\tau = 53.5$ for a simulation with $256^2$ spatial grid
points (broken line) and $512^2$ spatial grid points. These are the same simulations as
in Figures 3 and 4 respectively.

\bigskip

Figure 11. Log$_{10}V_1$ (left) and $\log_{10}V_2$ (right) for the line $u =
v$ for $u,v \in [0,2\pi]$ (vertical axis) and $\tau \in [10,59]$ (horizontal axis) are
shown on the same scale. The simulation of Figure 3 is used.

\bigskip

Figure 12. The final time step of Figure 11 is shown for $\log_{10}V_1$ (broken line),
$\log_{10}V_2$ (thick solid line), and $\varphi$ (thin solid line). The horizontal axis
is the line $u = v$ for $u,v \in [0,\pi]$
\vfill
\eject

\begin{figure}[bth]
\begin{center}
\makebox[4in]{\psfig{file=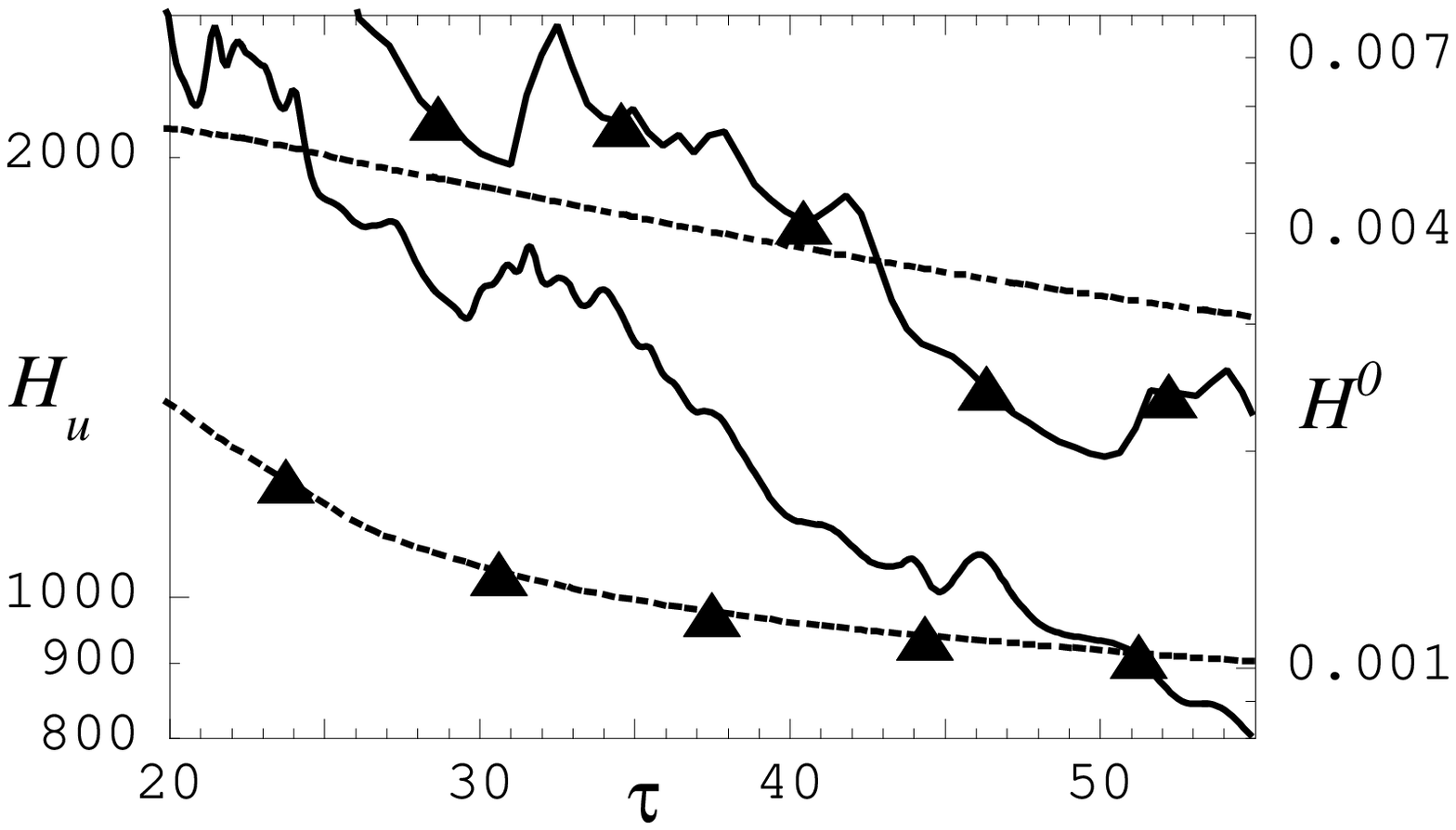,width=3.5in}}
\caption{}
\end{center}
\end{figure}

\begin{figure}[bth]
\begin{center}
\makebox[4in]{\psfig{file=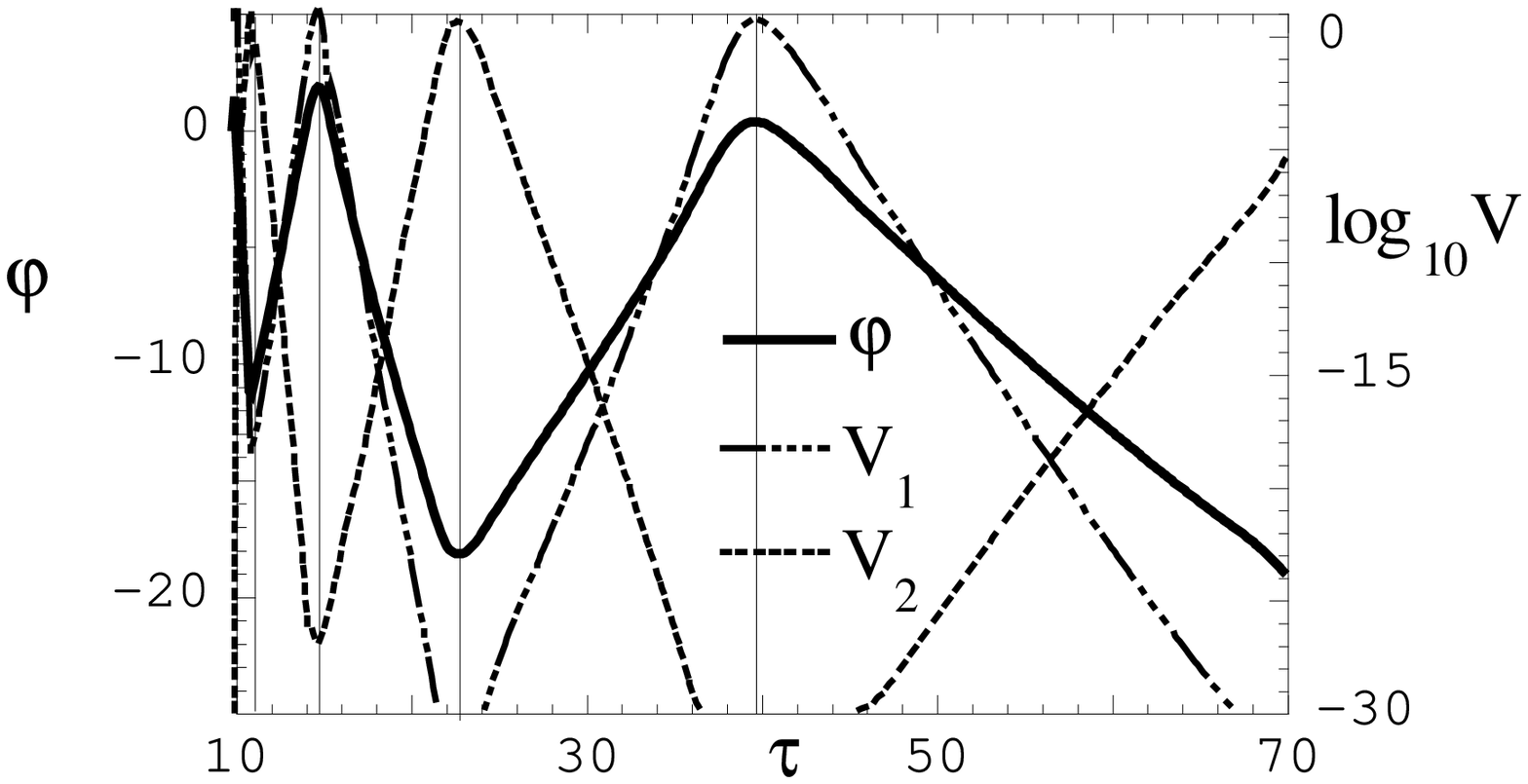,width=3.5in}}
\caption{}
\end{center}
\end{figure}

\begin{figure}[bth]
\begin{center}
\makebox[4in]{\psfig{file=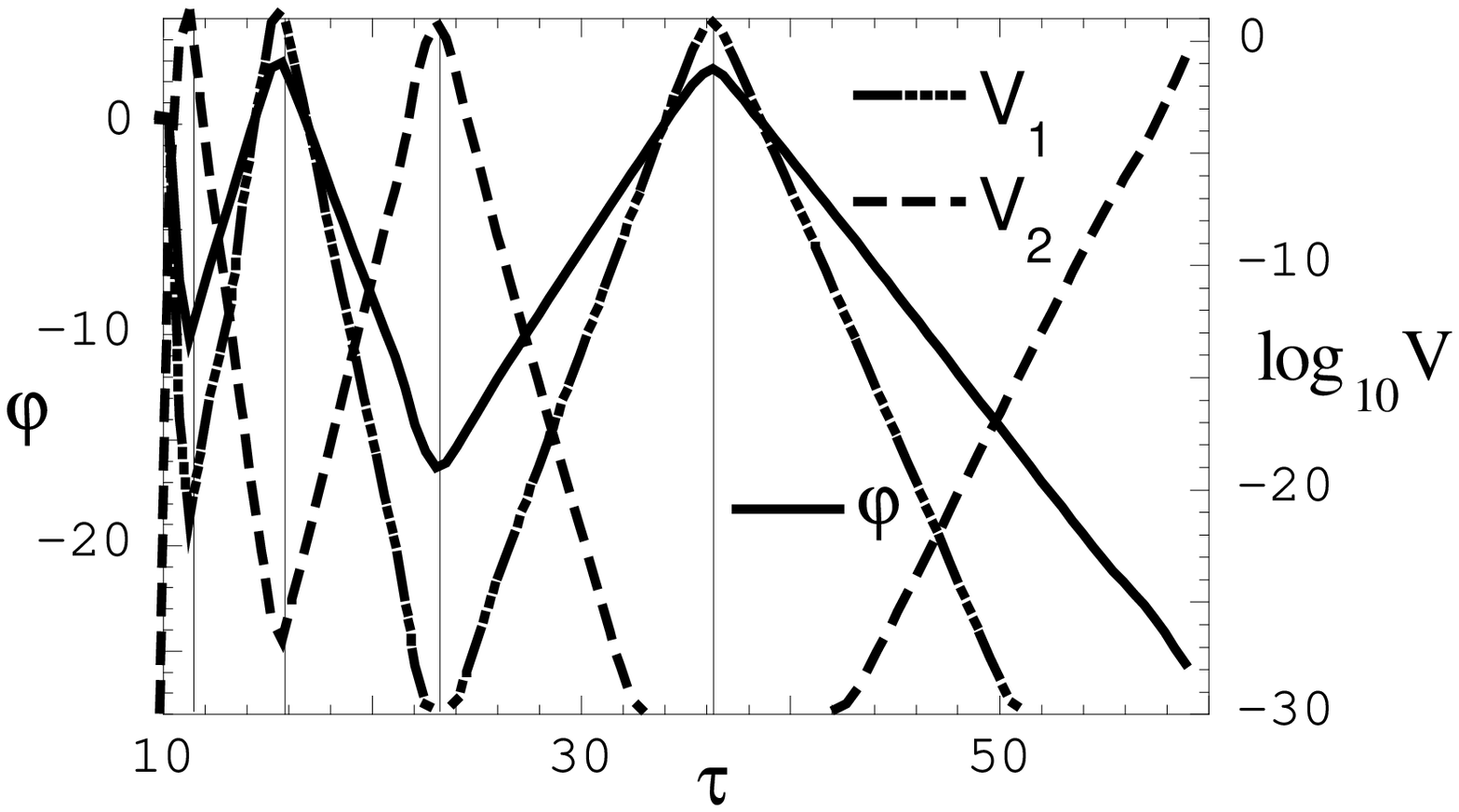,width=3.5in}}
\caption{}
\end{center}
\end{figure}
\vfill
\eject

\begin{figure}[bth]
\begin{center}
\makebox[4in]{\psfig{file=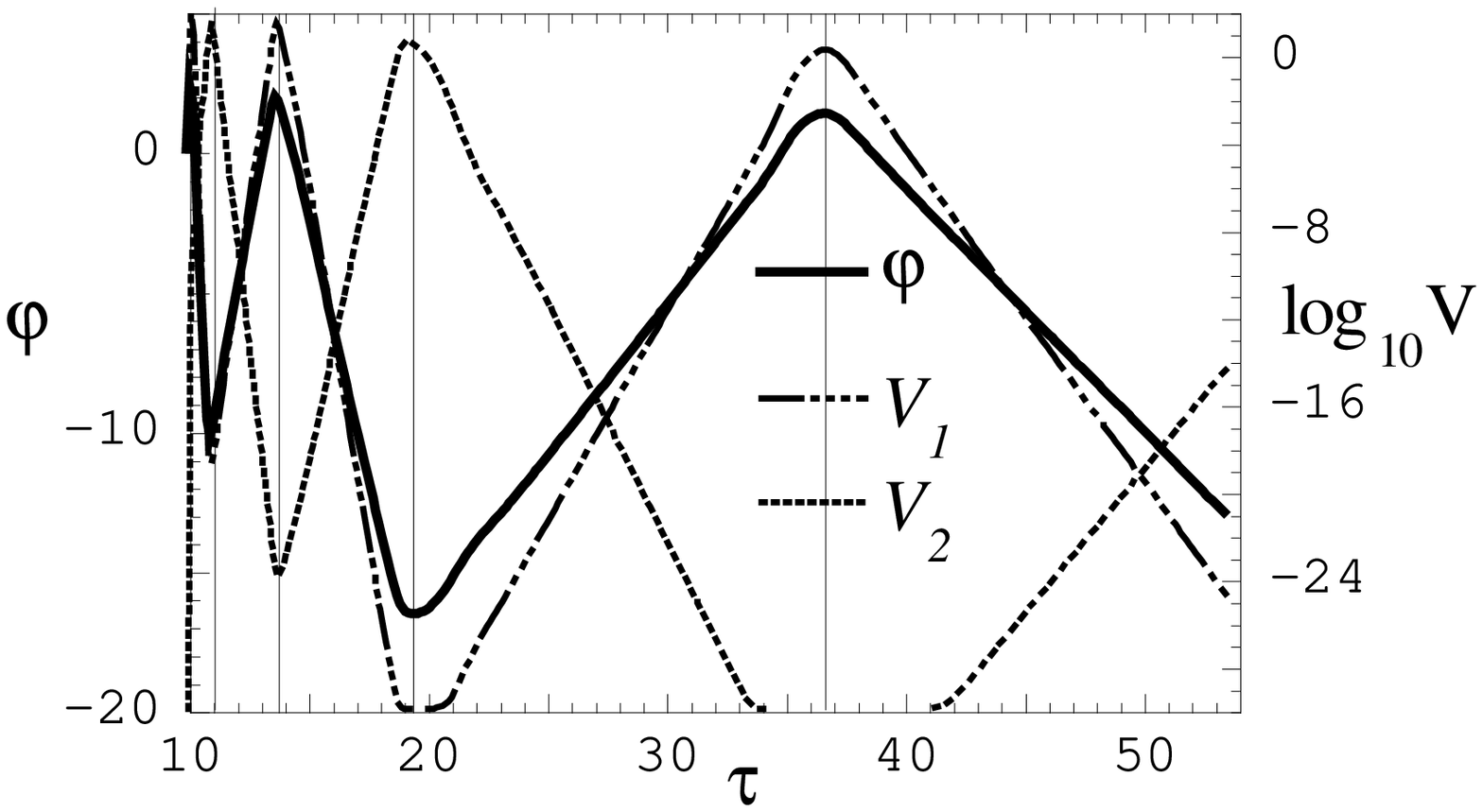,width=3.5in}}
\caption{}
\end{center}
\end{figure}

\begin{figure}[bth]
\begin{center}
\makebox[4in]{\psfig{file=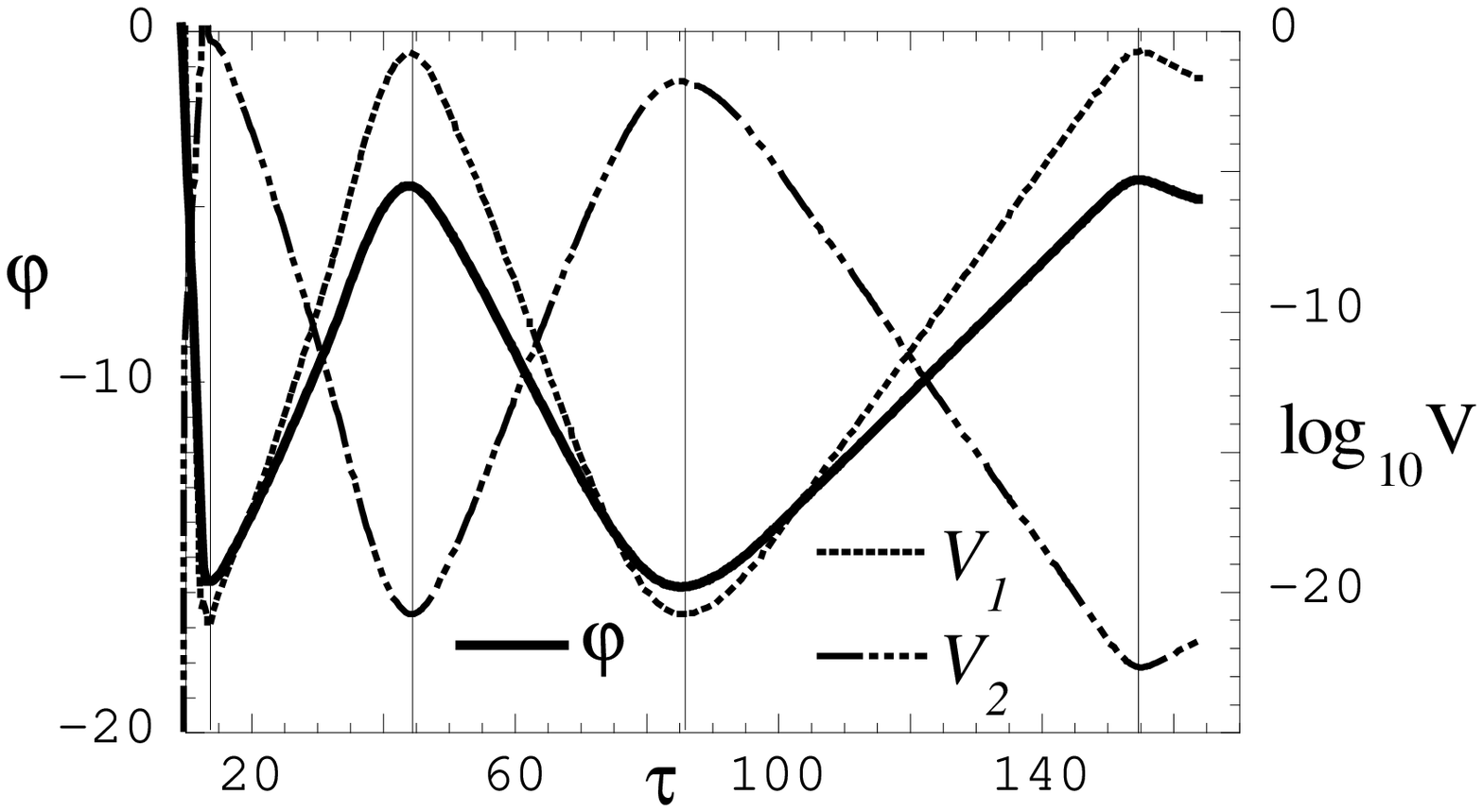,width=3.5in}}
\caption{}
\end{center}
\end{figure}

\begin{figure}[bth]
\begin{center}
\makebox[4in]{\psfig{file=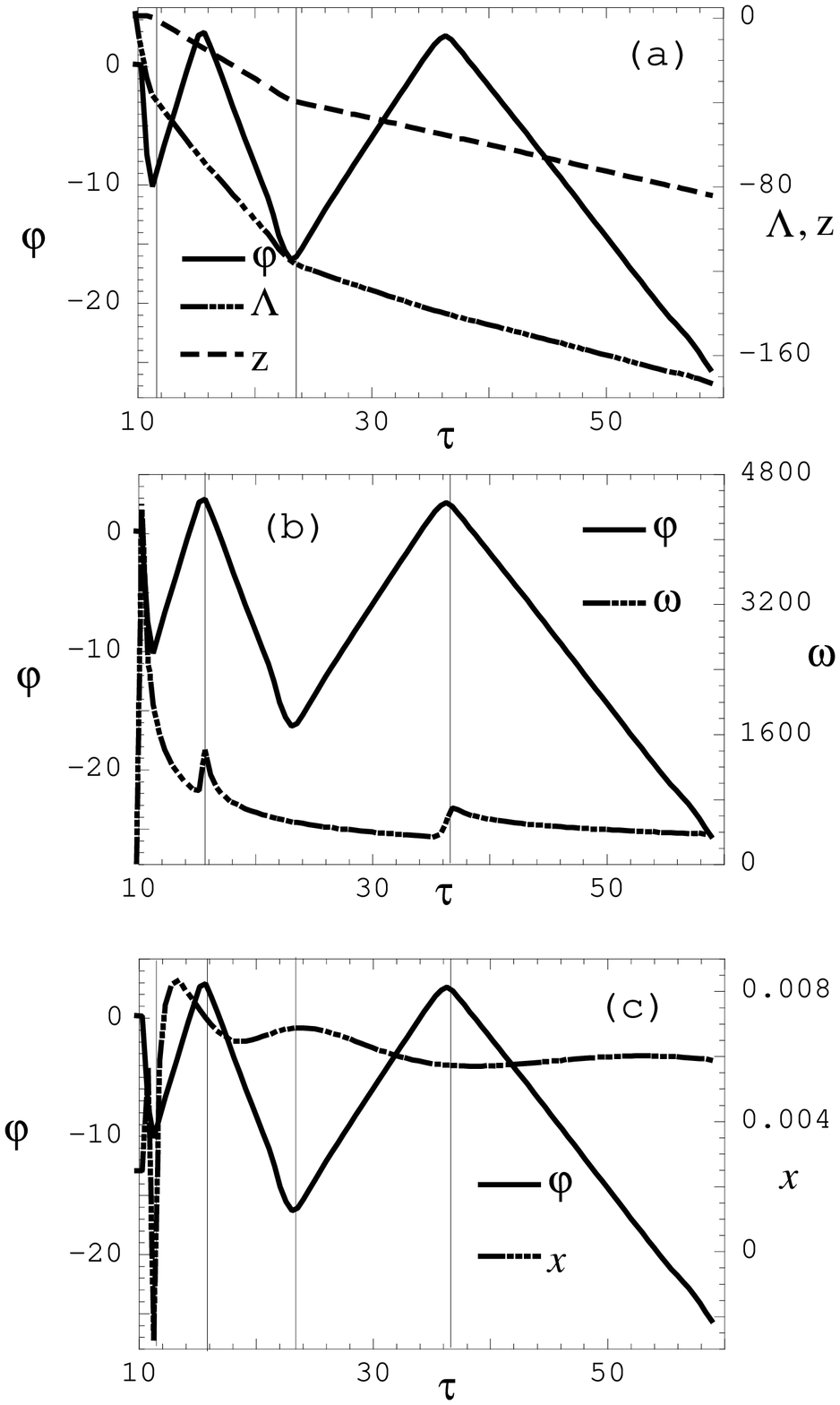,width=3.5in}}
\caption{}
\end{center}
\end{figure}

\begin{figure}[bth]
\begin{center}
\makebox[4in]{\psfig{file=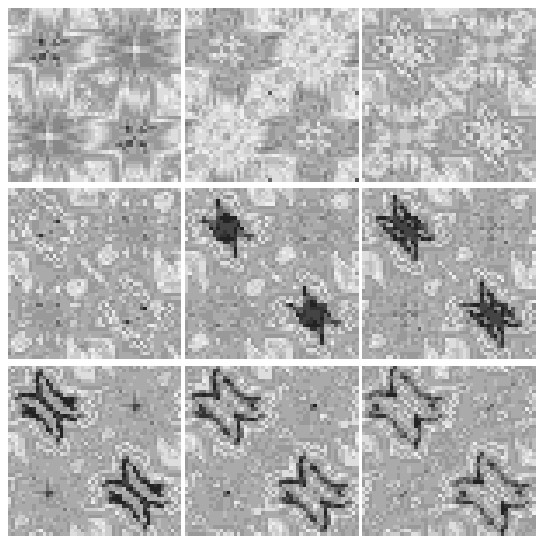,width=3.5in}}
\caption{}
\end{center}
\end{figure}

\begin{figure}[bth]
\begin{center}
\makebox[4in]{\psfig{file=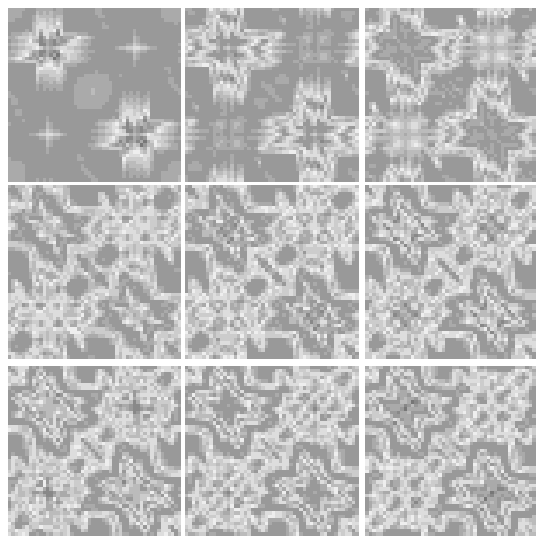,width=3.5in}}
\caption{}
\end{center}
\end{figure}

\begin{figure}[bth]
\begin{center}
\makebox[4in]{\psfig{file=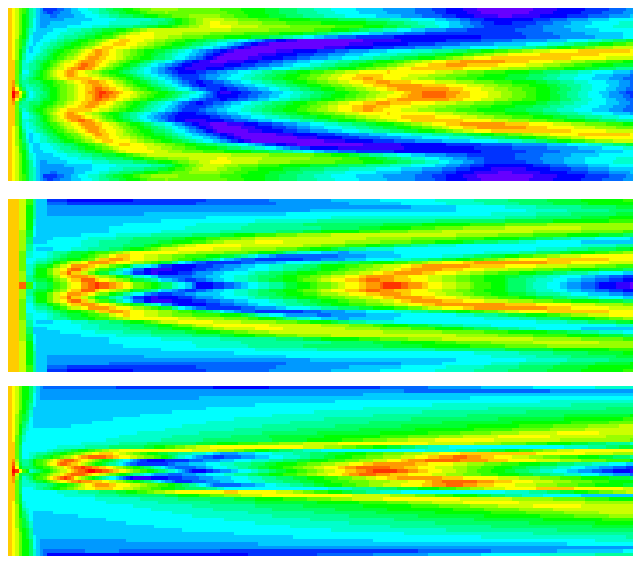,width=3.5in}}
\caption{}
\end{center}
\end{figure}

\begin{figure}[bth]
\begin{center}
\makebox[4in]{\psfig{file=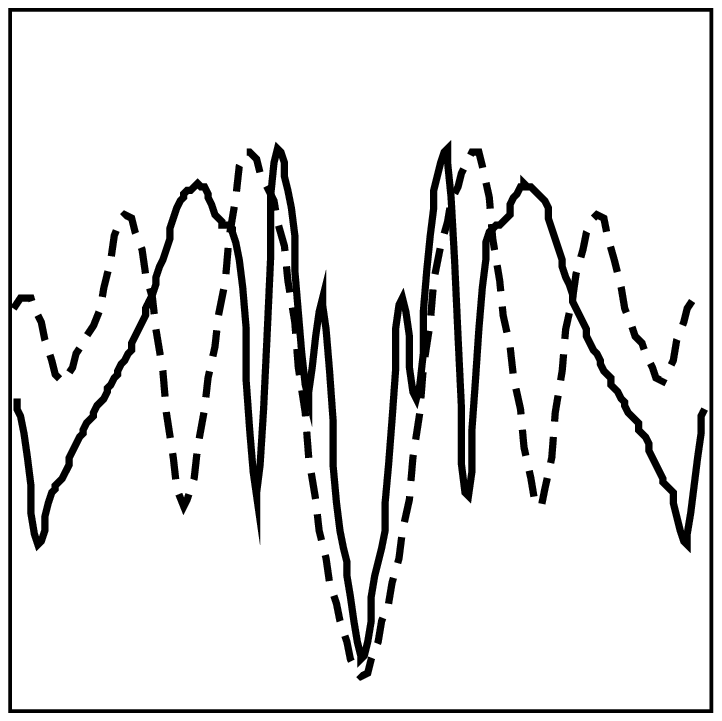,width=2in}}
\caption{}
\end{center}
\end{figure}

\begin{figure}[bth]
\begin{center}
\makebox[4in]{\psfig{file=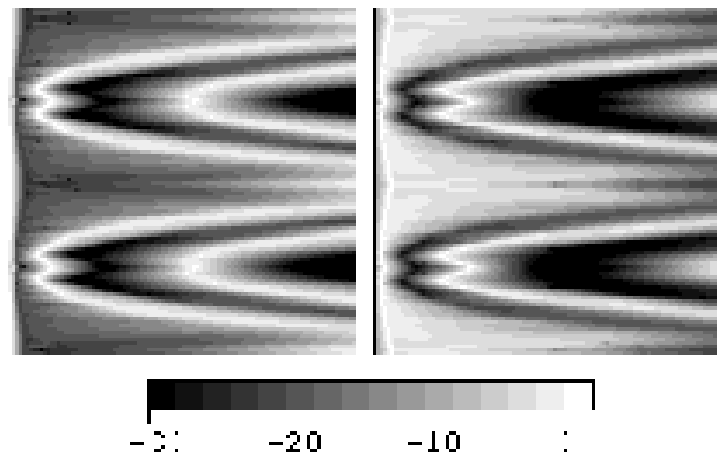,width=3.5in}}
\caption{}
\end{center}
\end{figure}

\begin{figure}[bth]
\begin{center}
\makebox[4in]{\psfig{file=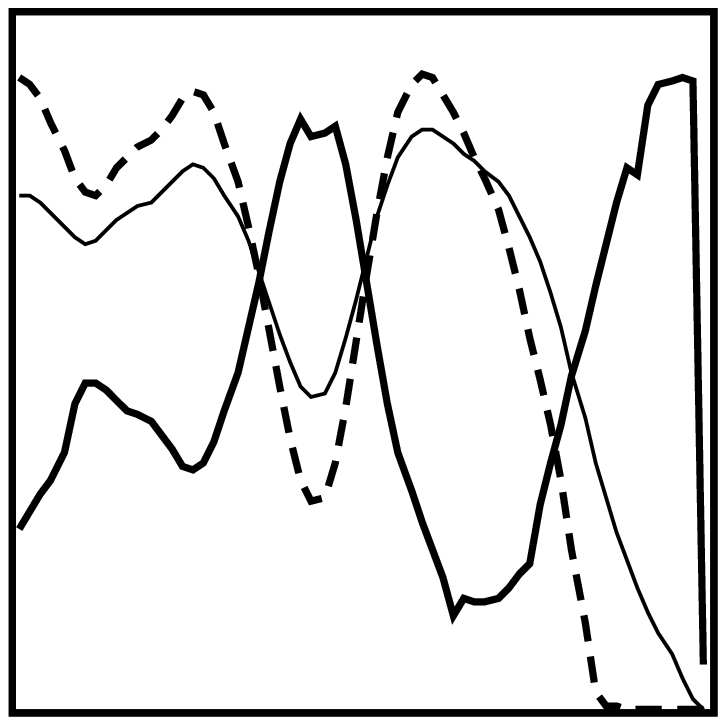,width=2in}}
\caption{}
\end{center}
\end{figure}


\begin{references}

\bibitem{penrose69}
{R. Penrose}, {Riv. Nuov. Cim.},
{\bf 1}, {252} (1969).

\bibitem{hawking67}
{S.W. Hawking}, {Proc. Roy. Soc. Lond. A}, {\bf 300}, {182} (1967).

\bibitem{hawking70}
{{S.W. Hawking} and {R. Penrose}}, {Proc. Roy. Soc. Lond. A},
{\bf 314},
{529} (1970).

\bibitem{hawking73}
{{S.W. Hawking} and {G.F.R. Ellis}},
{\em The Large Scale Structure of Space-Time},
(Cambridge University Press, Cambridge,1973).

\bibitem{wald84}
R.M. Wald, {\em General Relativity},
  (University of Chicago Press, Chicago,
 1984). 

\bibitem{belinskii69a}
{{V.A. Belinskii} and {I.M. Khalatnikov}},
{Sov. Phys. JETP},
{\bf 30},
{1174} (1969).

\bibitem{belinskii69b}
{{V.A. Belinskii} and {I.M. Khalatnikov}}, 
{Sov. Phys. JETP}, 
{\bf 29}, 
{911--917} (1969)

\bibitem{belinskii71a}
{{V.A. Belinskii} and {I.M. Khalatnikov}},
{Sov. Phys. JETP},
{\bf 32},
 {169--172} (1971).

\bibitem{belinskii71b}
{V.A. Belinskii}, {E.M. Lifshitz}, and {I.M. Khalatnikov}, Sov. Phys.
  Usp. {\bf 13},  745  (1971).

\bibitem{belinskii82}
{{V.A. Belinski}, {I.M. Khalatnikov} and {E.M. Lifshitz}},
{Adv. Phys.},
{\bf 31},
{639} (1982).

\bibitem{misner69}
C.~W. Misner, Phys. Rev. Lett. {\bf 22},  1071  (1969).

\bibitem{kasner25}
E. Kasner, {
  Trans. Am. Math. Soc.}, {\bf 27}, 155--162,
  (1925). 

\bibitem{chernoff83}
{{D.F. Chernoff} and {J.D. Barrow}},
{Phys. Rev. Lett.},
{\bf  50},
134 (1983).

\bibitem{barrow79}
{J.D. Barrow} and {F. Tipler}, Phys. Rep. {\bf 56},  372  (1979).

\bibitem{grubisic94a}
{B. Grubi\u{s}i\'{c}},in {\em Proceedings of the Cornelius Lanczos Symposium}, 
edited by {{J.D. Brown}, {M.T. Chu}, {D.C. Ellison}, and {R.J. Plemmons}}, ({SIAM},
{Philadelphia},1994).

\bibitem{isenberg90}
{J.A. Isenberg} and {V. Moncrief}, Ann. Phys. (N.Y.) {\bf 199},  84  (1990).

\bibitem{grubisic93}
{B. Grubi\u{s}i\'{c}} and {V. Moncrief}, Phys. Rev. D {\bf 47},  2371
  (1993).

\bibitem{berger93}
{B.K. Berger} and {V. Moncrief},
{Phys. Rev. D}, {\bf 48}, 4676 (1993).

\bibitem{eardley72}
{D. Eardley}, {E. Liang}, and {R. Sachs},
{J. Math. Phys.}, {\bf
13}, 99 (1972).

\bibitem{gowdy71}
R.H. Gowdy, {Phys. Rev. Lett.}, {\bf
  27}, 826 (1971). 

\bibitem{berger74}
B.K. Berger, {Ann. Phys. (N.Y.)}, {\bf 83}, 458, (1974).

\bibitem{berger97o}
B.K. Berger, P.T. Chru\'sciel, J. Isenberg, and V. Moncrief, Ann. Phys. (N.Y.), 
{\bf 260}, 117
(1997).

\bibitem{isenberg98}
J. Isenberg and S. Kichenassamy, ``Polarized $T^2$-symmetric vacuum spacetimes,''
preprint.

\bibitem{berger97b}
{B.K. Berger}, and {D. Garfinkle},
{Phys. Rev. D}, {\bf 57}, 4767 (1998). 

\bibitem{kichenassamy98}
S. Kichenassamy and A.D. Rendall, Class. Quantum Grav., {\bf 15}, 1339 (1998).

\bibitem{weaver98}
{M. Weaver}, {J. Isenberg}, and {B.K. Berger},
{Phys. Rev. Lett.}, {\bf 80}, 2980 (1998).

\bibitem{fujiwara93}
Y. Fujiwara, H. Ishihara, and H. Kodama, {Class. Quantum
Grav.}, {\bf 10}, 859 (1993).

\bibitem{leblanc95}
{V.G. LeBlanc}, {D. Kerr}, and {J. Wainwright},
{Class. Quantum Grav.},{\bf 12}, 513, (1995).

\bibitem{berger96a}
B.K. Berger,{Class. Quantum Grav.}, {\bf 13},
1273, (1996). 

\bibitem{moncrief86}
V. Moncrief, {Ann. Phys. (N.Y.)}, {\bf 167},
118, (1986). 

\bibitem{grubisic94}
{B. Grubi\u{s}i\'{c}} and {V. Moncrief}, Phys. Rev. D {\bf 49},  2792
  (1994).

\bibitem{berger97e}
{B.K. Berger} and {V. Moncrief},
{Phys. Rev. D}, to be published, gr-qc/9801078.

\bibitem{choptuik91}
M.W. Choptuik, Phys. Rev. D {\bf 44},  3124  (1991).

\bibitem{berger96c}
{B.K. Berger}, {D. Garfinkle}, and {E. Strasser},
{\em Class. Quantum Grav.}, {\bf 14},
L29 (1997). 

\bibitem{fleck76}
{J.A. Fleck}, {J.R. Morris}, and {M.D. Feit}, Appl. Phys. {\bf 10},  129
  (1976).

\bibitem{moncrief83}
V. Moncrief, Phys. Rev. D {\bf 28},  2485  (1983).

\bibitem{suzuki90}
M. Suzuki, Phys. Lett. A {\bf 146},  319  (1990).

\bibitem{suzuki91}
M. Suzuki, J. Math. Phys. {\bf 32},  400  (1991).

\bibitem{norton92}
A.H. Norton, {U}. of New South Wales preprint, 1992 (unpublished).

\bibitem{khalatnikov85}
{{I.M. Khalatnikov}, {E.M. Lifshitz}, {K.M. Khanin},
{L.N. Shchur}, and {Ya.G. Sinai}},
{J. Stat. Phys.},
{\bf 38},
97 (1985).

\bibitem{kirillov87}
{A.A. Kirillov}, and {A.A. Kochnev},
{JETP Lett.}, {\bf 46}, 435, (1987).

\end{references}
\end{document}